\documentclass{article}
\usepackage{spconf,amsmath,graphicx}
\usepackage{ bbold }
\usepackage{url}

\usepackage{enumitem}
\setlist{nosep, leftmargin=14pt}

\usepackage{mwe} % to get dummy images

\title{POST TRAINING UNCERTAINTY CALIBRATION OF DEEP NETWORKS\\ FOR MEDICAL IMAGE SEGMENTATION}
\name{Axel-Jan Rousseau$^{\star}$, Thijs Becker$^{\star}$, Jeroen Bertels$^{\dagger}$, Matthew B. Blaschko$^{\dagger, \ast}$, Dirk Valkenborg$^{\star, \ast}$}

\address{
$^{\star}$ I-Biostat, Data Science Institute, Hasselt University, Belgium\\
$^{\dagger}$ Processing Speech and Images, Department of Electrical Engineering, KU Leuven, Belgium\\
$^{\ast}$ shared last authorship
}
\begin{document}
%\ninept
%
\maketitle
\begin{abstract}
Neural networks for automated image segmentation are typically trained to achieve maximum accuracy, while less attention has been given to the calibration of their confidence scores. However, well-calibrated confidence scores provide valuable information towards the user. We investigate several post hoc calibration methods that are straightforward to implement, some of which are novel. They are compared to Monte Carlo (MC) dropout. They are applied to neural networks trained with cross-entropy (CE) and soft Dice (SD) losses on BraTS 2018 and ISLES 2018. Surprisingly, models trained on SD loss are not necessarily less calibrated than those trained on CE loss. In all cases, at least one post hoc method improves the calibration. There is limited consistency across the results, so we can't conclude on one method being superior.
In all cases, post hoc calibration is competitive with MC dropout.
Although average calibration improves compared to the base model, subject-level variance of the calibration remains similar.
\end{abstract}
\let\thefootnote\relax\footnote{This work has been submitted to the IEEE for possible publication. Copyright may be transferred without notice, after which this version may no longer be accessible.}
\section{Introduction}
\label{sec:intro}

Medical image segmentation is an invaluable aspect for many medical image analysis tasks, such as radiotherapy treatment planning, tumor volume measurement, analysis of tissue morphology, and others. %Fully connected
Convolutional neural
networks (NN), such as U-Net \cite{Ronneberger2015}, are being investigated to automate the segmentation process. Although the accuracy of segmentation models has improved, their ability to quantify their predictive uncertainty is often overlooked.
However, the ability of models to capture predictive confidence is a desirable property. This means that the model's (voxel-wise) predictions would directly reflect their classification accuracy. Calibrated confidences provide valuable extra information towards the end user and increase the trustworthiness of the model. For example, in a medical setting a low confidence could be a signal to hand over control to human doctors.

Dice and Jaccard metrics are among the most commonly used performance metrics for medical image segmentation. These metrics can be optimized directly by using their relaxations as loss functions, e.g.~soft Dice (SD) and soft Jaccard, leading to a better segmentation performance compared to the traditional use of a cross-entropy (CE) loss which optimizes voxel-wise accuracy \cite{Eelbode2020}. However, recent results indicate that models trained on SD do not produce well-calibrated models compared to CE loss \cite{BERTELS2021101833, Sander2018, Mehrtash2020}. In fact, SD optimization leads to hard predictions, even when there is true uncertainty present in the task, leaving no room for a direct interpretation as confidence scores \cite{BERTELS2021101833}. This irreducible uncertainty is called aleatoric uncertainty and is caused by noise, artifacts, or incomplete class labels.

Bayesian probability theory is used to reason about uncertainty, and Bayesian neural nets (BNN) \cite{neal2012bayesian} attempt to learn distributions over the model weights and allow for uncertainty estimation, albeit the training process is generally considered computationally expensive. Gal and Ghahramani showed that Monte Carlo (MC) dropout, where dropout is performed during inference, is an approximation of BNN \cite{Gal2016}. This method is one of the most popular for image segmentation, as it is simple to implement, at least if the original NN has dropout layers. Another approach is the use of deep ensembles \cite{Lakshminarayanan2016}. Deep ensembles train multiple networks from random initialization and use model averaging to obtain the uncertainty estimate. 
%MC dropout is related, in the sense that it can be interpreted as sampling different sub-NNs from the full NN. 
A clear downside of deep ensembles is the need to train multiple models from scratch. Networks can also be modified to learn to estimate the aleatoric uncertainty \cite{Kendall2017}. These networks return a probability distribution and have two outputs: a mean logit output for the prediction together with its variance.
Auxiliary networks trained on top of baseline models have also been proposed. DeVries and Taylor use such an auxiliary network to produce uncertainty maps starting from the baseline model's output to have an additional quality measure of the segmentation \cite{DeVries2018}.
Inspired by this, Jungo et al.~\cite{Jungo2020} used a simple auxiliary network consisting of three convolutional layers to predict voxel-wise segmentation errors from feature maps of the segmentation network.
Guo et al.~found that for classification tasks, modern deep NNs are not well calibrated \cite{Guo2017}. They evaluated the performance of various post hoc calibration methods on NN classifiers and found an extension of Platt scaling to be surprisingly effective. For segmentation tasks such post hoc calibration methods are almost never included, with some exceptions \cite{Karimi2019}.

In this paper, we investigate several post hoc calibration methods for segmentation, including Platt scaling, and compare them with MC dropout. We introduce simple auxiliary networks that can be applied to most segmentation methods and that improve calibration.
We furthermore show that if one takes a network trained with SD, its calibration can be significantly improved by fine-tuning the last layer with CE, with only a small drop in Dice score. Finally, we investigate the subject-level calibrations for both the base and calibrated networks.

\section{Methods}

\subsection{Performance metrics for calibration}
Consider a model with class predictions $\hat{y}$ and associated confidences $\hat{p}$. The confidence $\hat{p}$ is found by taking the logit output of the network $z$ through a sigmoid function: $\hat{p} = \sigma(z)$.
A model is perfectly calibrated when its class prediction $\hat{y}$ and associated confidence $\hat{p}$ are correct with a probability $p$:
\begin{equation}
P\left(\hat{y}=y | \hat{p}=p \right) = p,
\end{equation}
where $y$ is the true class. 
To measure how well-calibrated a model is, we can therefore define
\begin{equation}
E \left[ |P(\hat{y}=y | \hat{p}=p) - p| \right],
\end{equation}
where the average is taken over all predictions \cite{Guo2017}.
This equation cannot be calculated, but can be approximated with the expected calibration error (ECE) \cite{Naeini}. The predictions are binned into equally spaced bins $B_k$, $k \in [1, K]$ (we use $K=20$). The accuracy of bin $B_k$ is defined as $\text{acc}(B_k) = \sum_{i\in B_k} \mathbb{1}(\hat{y_i}=y_i) / |B_k|$ and its confidence as $\text{conf}(B_k) = \sum_{i\in B_k} \hat{p_i} / |B_k|$, where $|B_k|$ is the number of elements in the bin.
% and for each bin,  the average confidence and accuracy is calculated. 
% The accuracy of bin $B_k$ is defined as
% \begin{equation}
%     \text{ACC}(B_k) = \frac{1}{|B_k|}\sum_{i\in B_k} \mathbb{1}(\hat{y_i}=y_i),
% \end{equation}
% and the average confidence of the bin is
% \begin{equation}
%     \text{CONF}(B_k) = \frac{1}{|B_k|}\sum_{i\in B_k} \hat{p_i},
% \end{equation}
% where $|B_k|$ is the number of elements in the bin.
The ECE is the weighted average of the difference in 
accuracy and confidence of the bins:
\begin{equation}
\text{ECE} = \sum_{k=1}^{K}\frac{|B_k|}{N}| \text{acc}(B_k)-\text{conf}(B_k)|,
\end{equation} 
% \begin{equation}
% \text{ECE} = \frac{1}{N} \sum_{k=1}^{K} \left| \sum_{i\in B_k} \left( \mathbb{1}(\hat{y_i}=y_i) - \hat{p_i} \right) \right|,
% \end{equation} 
with $N$ is the total number of elements in all bins.
This can be represented visually with reliability diagrams. These plot the accuracy of the bins against their confidence. A reliability diagram for a perfectly calibrated model plots the identity function.
% Because of the bin-size weighting, the often highly-confident and accurate background pixels have a large effect on the ECE. We therefore only consider those voxels belonging to the brain for the ECE calculation, similar to \cite{Jungo2020}.

\label{sec:calibration}
\subsection{Calibration methods}
We investigate the following four calibration methods. Source code is available at \url{github.com/AxelJanRousseau/PostTrainCalibration}
% Source code available at \url{https://anonymized.url/review}. 

\textbf{Platt scaling:} A simple parametric method that is often used for classification tasks \cite{Platt1999}. It involves fitting a logistic regression model on top of the logit outputs $z$ of the network, whilst keeping the model's weights fixed. For a segmentation task, this is equivalent to training a convolutional layer with a single kernel of size 1 on top of the logit outputs of the network, resulting in only 2 extra weights that need to be optimized using CE. 

\textbf{Auxiliary network:} Keeping in line with this convolution analogy, we also tested single convolutional layers using larger kernel sizes. These can be seen as a simple, shallow auxiliary network, or as a generalized version of Platt scaling, where the scaling is influenced by neighbouring voxels. Kernel sizes of $k$ = 5, 7, 9 and 11 were tested, but sizes larger than 5 provided little benefit. Therefore all reported values are for $k = 5$.

\textbf{Fine-tuning:} Instead of adding extra layers, we consider the effect of fine-tuning the last learnable layer of the network. Starting from the base network, the weights of all layers except the last one are frozen, and training is resumed using CE loss. As SD loss is expected to lead to poor calibration, we expect that fine-tuning with CE can improve the calibration of models trained using SD.

\textbf{MC Dropout:} 
Dropout is enabled during test time and multiple predictions are sampled (in this case 20). We either retrain the base model with dropout layers included, or we add dropout layers at test time without retraining.

\subsection{Datasets}
We tested the calibration performance on the publicly available training datasets of BraTS 2018 \cite{Bakas2018,Bakas2017,Menze2015} (BR18) and ISLES 2018 \cite{Maier2017} (IS18). The BR18 dataset contains multi-modal MRI scans of 285 patients with brain tumors. For each patient T1-weighted (T1), post-contrast T1-weighted (T1c), T2-weighted (T2) and T2 Fluid Attenuated Inversion Recovery (FLAIR) scans were provided along with manual consensus delineations of whole tumor (WT), active tumor (AT) and tumor core (TC) tissue. Here we only consider the WT segmentation task.
The IS18 dataset contains CT perfusion scans of 94 cases with an acute ischemic stroke. The available perfusion parameter maps are cerebral blood volume (CBV), cerebral blood flow (CBF), mean transit time (MTT), time to peak of the residue function (Tmax) and the raw CT perfusion data.
The input volumes were resampled to a 2~mm isotropic voxel size and center patches of size 136$\times$136$\times$82 were used. All available modalities were used  and concatenated along the feature dimension as input to the network, except for raw perfusion data for IS18, which was neglected.
Because of the bin-size weighting in the ECE metric, the often highly-confident and accurate background pixels have a large effect. We therefore only consider those voxels belonging to the brain for the ECE calculation, similar to \cite{Jungo2020}.

\subsection{U-Net architecture and training}
The network architecture used is the modified 3D U-Net-like network from \cite{Eelbode2020}. 
To speed up convergence the model was pre-trained using CE loss, after which training continued using SD or CE loss. Weights for these models are indicated by CE-SD and CE, respectively. In addition, models were also trained directly on SD loss. We used the same 5-fold cross-validation and training scheme as in \cite{Eelbode2020}. No data-augmentation was used for training of the calibration methods.

Training of Platt scaling and the auxiliary networks was done on batches of 64 z-slices taken from the 3D volumes. The Adam optimizer was used with an initial learning rate of $5\cdot10^{-3}$ to optimize the CE. The learning rate decreased with a factor 10 when the loss reached a plateau. Training continued for a maximum of 50 epochs, or until the loss on the validation fold stopped improving.

For fine-tuning, training was done on batches of 2 3D volumes. The Adam optimizer, loss function, learning rate schedule, and early stopping are used as before, but the initial learning rate was $10^{-4}$ for models trained with CE loss and $10^{-3}$ for models trained on SD loss.

Because initially no dropout layers were used to train the base models, they were inserted to the base models later. We then retrain all layers after the inserted dropout layers to ensure the dropout does not wreck the input distributions of the subsequent layers. Two dropout settings used in previous literature are considered \cite{Kendall2017a}: In the first setting, referred to as MC-Decoder, dropout is added before each decoder block of the network and before the last convolutional output. The second setting, referred to as MC-Center, applies dropout after the lowest two encoder blocks and before the lowest two decoder blocks. Retraining of the layers after dropout was done with SD for models using SD-based model weights, and CE for models using the CE-based model weights. There was no single learning rate setting that produced good results across dataset, weight or dropout settings. We therefore trained each model using three different initial learning rate settings of $10^{-3}$, $10^{-4}$ and $10^{-5}$ and report on the best performing model. A learning rate schedule together with early stopping were used. Not training after adding dropout typically lowered calibration compared to the base model. Training always resulted in better performance.

\section{Results and discussion}

\begin{table}[htb]
\setlength{\tabcolsep}{2pt}
\caption{Results on BR18. Values in bold indicate the best result, or not significantly different from the best result (Wilcoxon signed-rank test $p<0.05$).}
\label{tab:brats-table}
\begin{tabular}{c|ccc|ccc}
\textit{weights\(\rightarrow\)} & CE  & CE-SD     & SD &   CE  & CE-SD     & SD   \\ \hline
method      &       &   Dice    &   &  & ECE\%\\ \hline
base model  & 0.848  &  \textbf{0.870}    & \textbf{0.878}  &  1.765 &  1.429   & 1.367   \\
Platt       & 0.851  &  \textbf{0.869}    & 0.877 & 1.246 &  1.186  & 1.087   \\
auxiliary   & 0.851  &  \textbf{0.869}    & 0.877 &  1.224 &  1.158   & 1.067  \\
fine-tune   & 0.851  &  0.869    & 0.876  & 1.269 &  \textbf{1.141}   & \textbf{1.028}   \\
MC-Decoders & 0.854  &  0.863    & 0.873  & 1.231 &  1.323  & 1.247    \\
MC-Center & \textbf{0.855} & \textbf{0.867}    & 0.875  &  \textbf{1.180} & 1.408   & 1.308      \\ \hline
 
\end{tabular}
\end{table}

\begin{table}[htb]
\setlength{\tabcolsep}{2pt}
\caption{Results on IS18. Values in bold indicate the best result, or not significantly different from the best result (Wilcoxon signed-rank test $p<0.05$).}
\label{tab:isles-table}
\begin{tabular}{c|ccc|ccc}
\textit{weights\(\rightarrow\)} & CE  & CE-SD     & SD &   CE  & CE-SD     & SD   \\ \hline
method      &       &   Dice    &   &   & ECE\%\\ \hline
base model    & 0.462      &  \textbf{0.538}    & \textbf{0.527} & 1.902     &  2.765   & 2.981     \\
Platt           & 0.200      &  0.499    & 0.407     & 4.173     &  1.917   &2.248      \\
auxiliary       & 0.451      &  0.509    & 0.466     & 1.800     &  2.076   & \textbf{2.116}      \\
fine-tune      & \textbf{0.467}      &  0.523      & 0.491     & 1.808     &  \textbf{1.822}   & 2.191      \\
MC-Decoder      & 0.463      &  0.515    & 0.498     & 1.711     &  2.352   & 2.312       \\
MC-Center       & 0.455      &  0.519    & 0.497     & \textbf{1.694}     &  1.985   & 2.305 \\ \hline

\end{tabular}
\end{table}

We calculated the mean of the per-volume Dice and ECE of every subject in the validation folds. 
Tables \ref{tab:brats-table} and \ref{tab:isles-table} show the results for the base model and the four calibration methods on the BR18 and IS18 datasets. 
As expected, the results show a superior segmentation performance in terms of Dice score for SD-based methods. Surprisingly, on BR18 the calibration of CE-based models is not better compared to SD-based models. 

On BR18, all methods improve the ECE over the baseline without appreciably lowering the Dice score. In fact, there is a slight increase in the Dice score for the CE model.
Fine-tuning turns out to be superior for SD-based models, although the results are close to both Platt scaling and auxiliary networks. The MC-Center method turns out to be superior for CE-based models for both Dice and ECE. However, MC-Center has the highest ECE for all calibration methods for the CE-SD and SD model. The MC-Decoders method consistently scores amongst the worst calibration methods on BR18.

\begin{figure}[htb]

\begin{minipage}[b]{1.\linewidth}
  \centering
  \centerline{\includegraphics[width=1 \columnwidth]{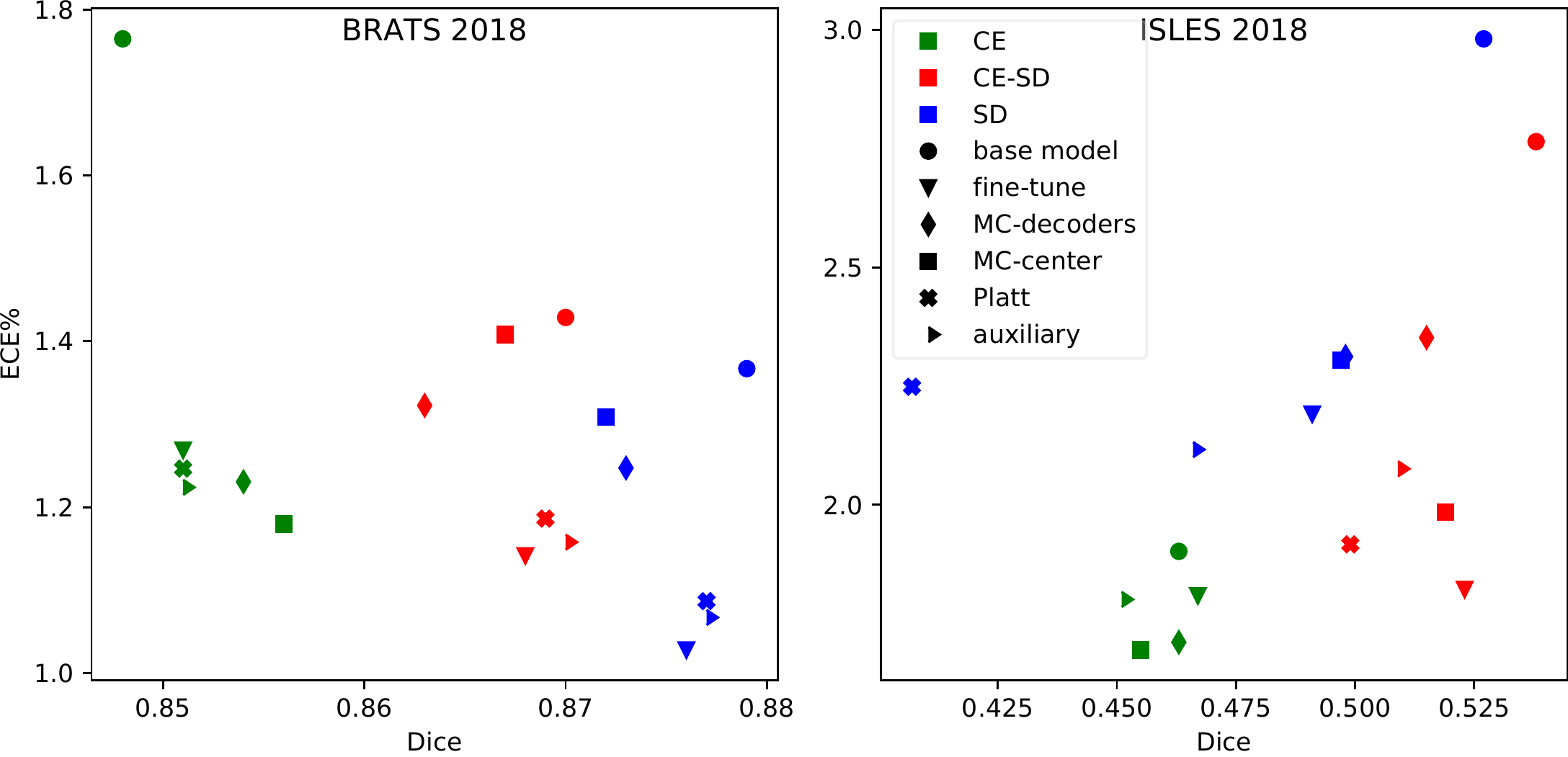}}
%  \vspace{2.0cm}
%  \centerline{(a) Result 1}\medskip
\end{minipage}
\caption{Scatter plots of ECE\% versus Dice score.}
\label{fig:scatter}
\end{figure}

Figure \ref{fig:scatter} shows scatter plots of ECE against Dice. For the SD trained weights we observe that the MC methods are Pareto-dominated (ECE and Dice can be improved by choosing another method), and fine-tuning and auxiliary networks are both Pareto efficient. Only SD trained models are on the Pareto frontier.

Results on IS18 show a different picture. The segmentation of ischemic stroke lesions from CT data is more challenging as MRI is considered to be the gold standard for lesion annotation. This means there is significant aleatoric uncertainty present in the IS18 dataset. All but one calibration methods lower the Dice score significantly, with Platt scaling having the largest negative impact. If each network is considered separately, fine-tuning is always Pareto efficient as can be seen on the scatterplot in Figure \ref{fig:scatter}. 
In contrast to the results on BR18, no networks trained with SD are on the Pareto frontier.
For the CE trained weights, the MC methods performs well: both are Pareto optimal on IS18.

Overall, we can conclude that simple post hoc methods have similar performance compared to MC dropout. Models trained on CE seem more suited for MC dropout. In contrast to previous results \cite{BERTELS2021101833, Sander2018, Mehrtash2020}, we observe that networks trained with SD can be surprisingly well-calibrated, as observed on BR18. On IS18, the calibration of networks trained with SD is significantly worse. This highlights the strong dependency on the dataset for calibration results.

\begin{figure}[ht]
\begin{minipage}[b]{1\linewidth}
    \centering
    \centerline{\includegraphics[width=1 \columnwidth]{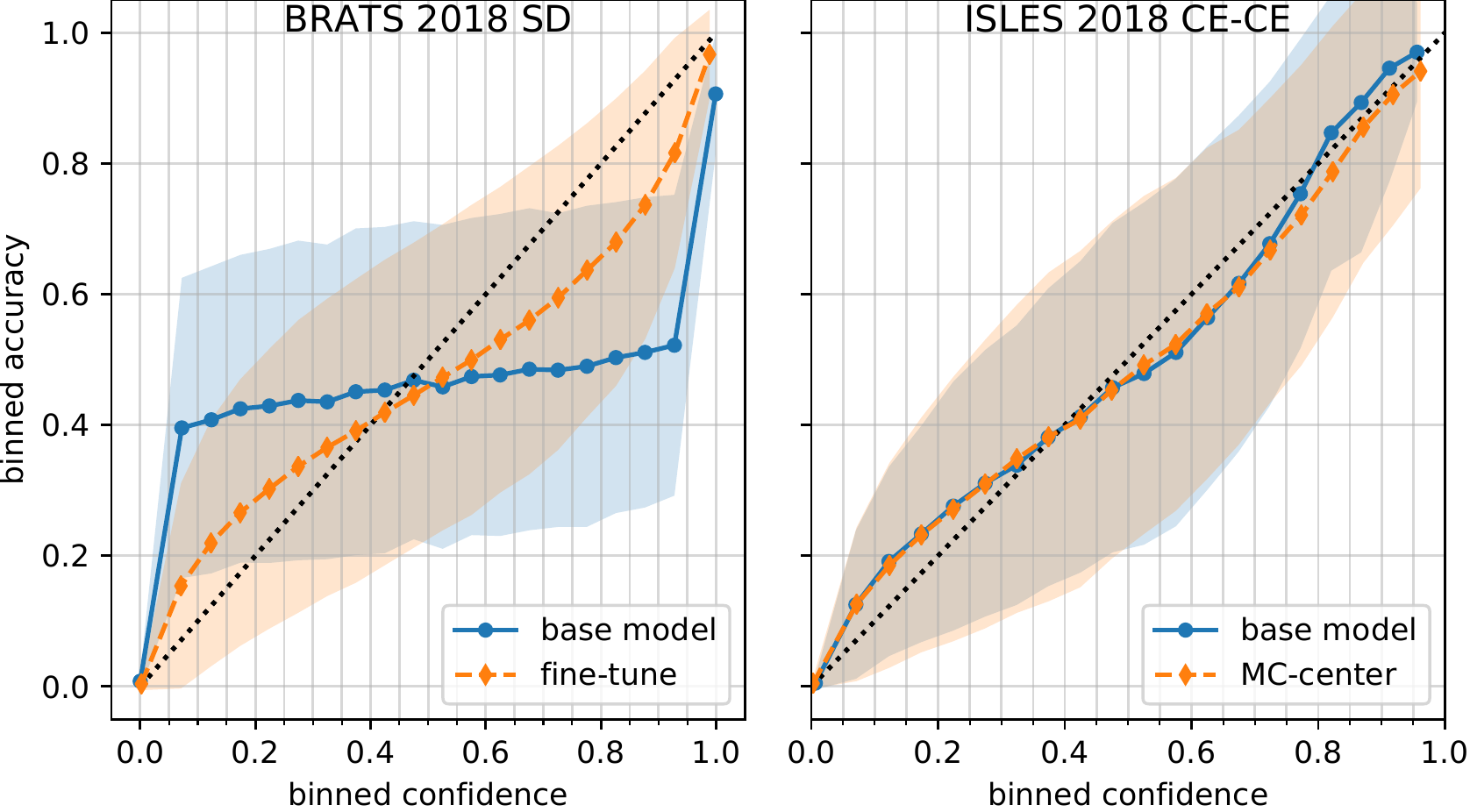}}
\end{minipage}

\caption{Reliability diagrams. For each bin the accuracy (points) and standard deviation of the volumes (shaded area) is shown.}
\label{fig:calibration_plots}
\end{figure}

\begin{figure}[ht]
\begin{minipage}[b]{1\linewidth}
    \centering
    \centerline{\includegraphics[width=1 \columnwidth]{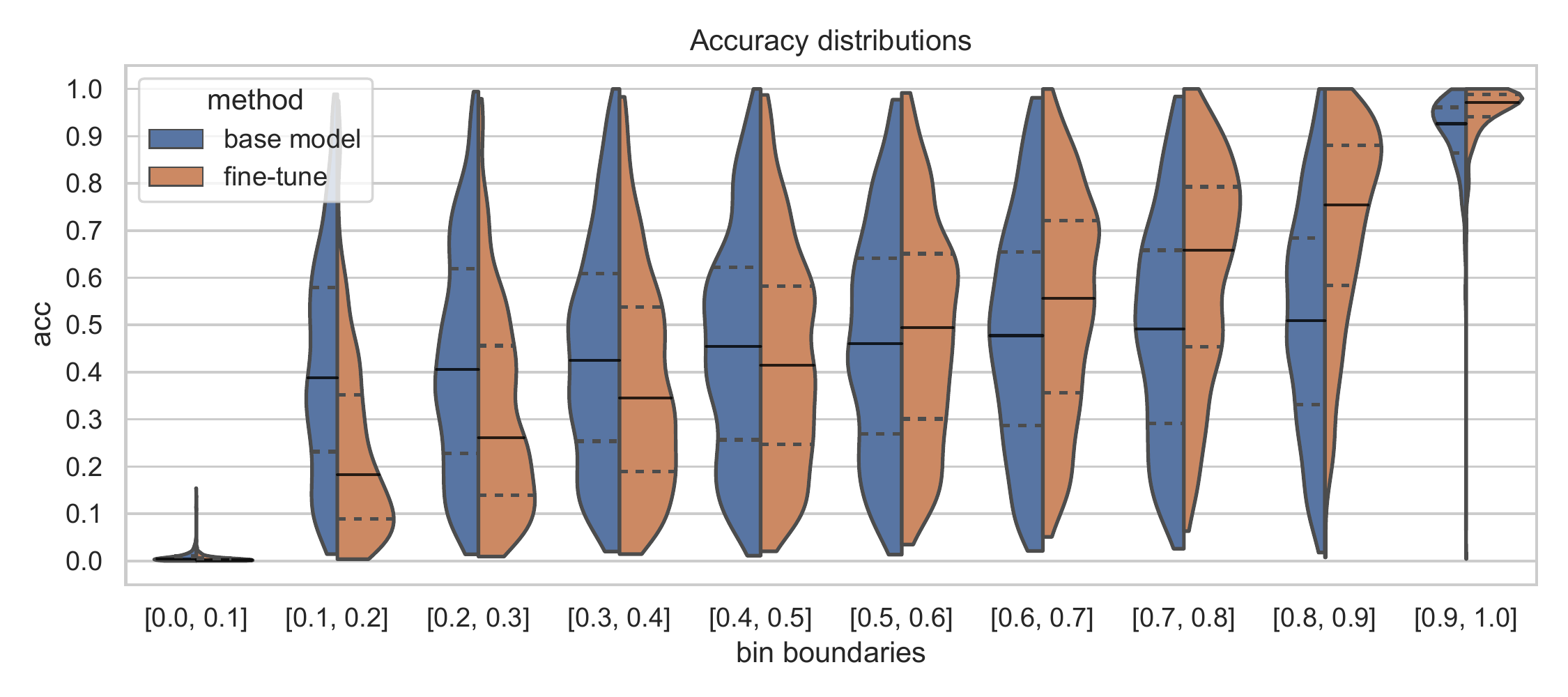}}
\caption{Violin plots of the subject-level accuracy distributions of the bins, for the BR18 SD model. Median (full lines) and first and third quartiles (dashed lines) are shown.}
\label{fig:violin_plots}
\end{minipage}
\end{figure}

It was noted by Jungo et al.~\cite{Jungo2020} that good average calibration doesn't necessarily translate to good subject-level calibrations. Although their model was well-calibrated on average, a significant number of subjects were under- or overconfident.
We examined the distribution of the accuracy in each bin and find that their standard deviations and interquartile ranges do not change much after calibration. The calibration plots in figure \ref{fig:calibration_plots} show that the difference in mean accuracy and confidence of each bin decreases, and as such the plot moves closer to the ideal line after calibration for SD weights. With CE weights, this shift is not apparent on the plot. As the standard deviations of the bins stays the same, we can conclude that calibration has no adverse effects on the amount of under- or overconfident subjects.
We further investigate the subject-level quality by looking at the subject-level distribution over several bins in the calibration curve, as shown in figure \ref{fig:violin_plots}. Despite improving on average, the amount of outliers doesn't considerably change.

\section{Conclusion}
\label{sec:Conclusion}
Neural networks for segmentation are often trained to achieve optimal performance on metrics such as Dice and Jaccard, while less attention is given to their uncertainty calibration. 
%Calibrated models can provide helpful information for human-in-the-loop settings.
We investigated post hoc methods that calibrate models without needing to retrain them. 
There are large differences in calibration performance depending on the segmentation task. 
%On the more challenging IS18 data the calibration methods had a larger negative impact on the dice scores compared to BR18. 
% There is limited consistency across the results, and 
% There is no clear best calibration method and a strong depe. 
We find that simple post hoc methods are competitive with the popular MC dropout. MC dropout is most efficient for models trained with cross-entropy. Surprisingly, models trained with soft Dice are better calibrated than those trained with cross-entropy for BraTS 2018. 
Consistent with previous results \cite{Jungo2020}, we find that despite an improvement in average calibration performance, several subjects remain under- or overconfident. Using segmentation maps as feedback for uncertainty should therefore be done with caution.

\subsection*{Acknowledgments}
%\label{sec:acknowledgments}
This research received funding from the Flemish Government under the “Onderzoeksprogramma Artificiële Intelligentie (AI) Vlaanderen” programme.
The computational resources and services used in this work were provided by the VSC (Flemish Supercomputer Center), funded by the Research Foundation - Flanders (FWO) and the Flemish Government - department EWI.
The authors have no relevant financial or non-financial interests to disclose.

\subsection*{Compliance with Ethical Standards}

This research study was conducted retrospectively using human subject data made available in open access at \url{https://ipp.cbica.upenn.edu/} for BraTS 2018 and \url{https://www.smir.ch/ISLES/Start2018} for ISLES 2018. Ethical approval was not required as confirmed by the license attached with the open access data.
\bibliographystyle{IEEEbib}
\bibliography{Calibration}

\end{document}